\documentclass[12pt]{iopart}

\usepackage{iopams,graphicx}  
\begin{document}

\title{Zero-energy states in rotating trapped Bose-Einstein condensates}

\author{Tapio Simula}

\address{School of Physics, Monash University, Victoria 3800, Australia}
\ead{tapio.simula@monash.edu}
\begin{abstract}
We have calculated low-lying quasiparticle excitation spectra of rotating three-dimensional Bose-Einstein condensates. As opposed to the prediction of hydrodynamic continuum theories, we find a minimum in the Tkachenko mode spectrum at intermediate rotation frequencies of the harmonic trap. Such a minimum can harbour a Tkachenko quasiparticle with zero excitation energy. We discuss the experimental signatures of such a zero mode. 
\end{abstract}

\maketitle

\section{Introduction}
The significance of zero-energy excitations in quantum field theories of both elementary and quasiparticles has substantially increased since Majorana predicted the appearance of a massless particle, which acts as its own anti-particle \cite{Majorana1937a}. Neutrinos have been suggested as candidates for Majorana fermions and more recently, non-elementary Majorana quasiparticles have sparked major interest due to the prospects of harnessing them for fault tolerant topological quantum computing \cite{Kitaev1997a,Kitaev2001a,Nayak2008a}. In such quantum information processing protocols quantized vortices of non-Abelian character could be moved adiabatically \cite{Virtanen2001aa} to achieve braiding of topological charges. This could be facilitated by Majorana quasiparticle zero modes, which are predicted to appear trapped inside vortex cores in chiral p-wave paired superconductors and protected by coupling to edge modes \cite{Volovik1999a,Read2000a,Nayak2008a}. Signatures of Majorana zero modes in superconductor-semiconductor nanowires have recently been observed \cite{Mourik2012a}. 

Zero-energy excitations, often called Nambu--Goldstone bosons, have been central to discussions of Lorentz-invariant field theories with spontaneously broken continuous symmetries \cite{Nambu1960a,Nambu1961a,Goldstone1961a,Goldstone1962a,Watanabe2012a}. The mass of elementary particles is thought to emerge due to shifts in the energies of such zero modes \cite{Stueckelberg1938a,Schwinger1962a,Anderson1963a,Higgs1964a,Englert1964a,Guralnik1964a,Aad2012a,Chatrchyan2012a}. Nambu--Goldstone modes also emerge in systems without Lorentz invariance. In condensed matter systems, phase transitions involving spontaneous symmetry breaking, such as the Bose--Einstein condensation emerge with associated Nambu--Goldstone bosons. In superconductors, however, such a zero mode is known to be gapped due to the Coulomb interaction \cite{Anderson1958a,Anderson1958b,Anderson1963a}. Goldstone theorem predicts the emergence of a zero energy (quasi)particle corresponding to every continuous symmetry of the Hamiltonian absent in the ground state of the system \cite{Goldstone1961a,Goldstone1962a}, and such symmetry absentees are therefore said to be spontaneously broken. However a direct application of this principle to non-relativistic systems can lead to seeming contradictions. There remains some confusion in particular regarding the number of zero modes that should be observed in such systems. For example, it may come as a surprise that for atoms in a crystal lattice, a state which clearly breaks six continuous symmetries (three translational plus three rotational), there only exists three (two transverse and one longitudinal) gapless phonon modes. It has been suggested that the Nambu--Goldstone modes associated with the rotational symmetry breakings are absent in atomic crystals because of the linear relationship between Noether charge densities for translation and rotation, making the rotational zero modes redundant \cite{Watanabe2012a}.
   
Similar situation prevails in the harmonically trapped rotating Bose--Einstein condensates. It is well known that due to the process of Bose--Einstein condensation the condensate ground state establishes a spatial phase coherence. Consequently, the Bogoliubov quasiparticle excitation spectrum becomes gapless acquiring a zero energy excitation mode. This is the Nambu--Goldstone (quasiparticle) boson also known as the Bose--Einstein condensate. When such condensate is rotated the ground state changes its topology by nucleating a quantized vortex crystal lattice. Similarly to atoms in solids, such vortex lattice breaks six additional continuous symmetries but experiments have not detected any emergent zero modes associated with the formation of the vortex lattice. A solution to this puzzling feature is suggested in \cite{Watanabe2012a} by noting that already the nonrotating condensate ground state breaks Galilean invariance and therefore the six additional symmetries broken by the vortex crystal are redundant and already catered for by the zero mode associated with the formation of the condensate. Notice also that the external harmonic trapping of the condensate also breaks the translation invariance of the condensate ground state and as such should be treated on an equal footing with other translation symmetry breaking fields. Indeed, the continuous translation symmetries absent in the vortex lattice state are already broken in the nonrotating ground state by the external trapping potentials. The fact that in a rotating harmonically trapped Bose--Einstein condensate there is (normally) only one zero mode can also be argued from dynamical perspective. Since the motion of the condensate atoms and the vortex lattice do not constitute independent degrees of freedom they cannot usually be treated as independent fields, which would possess their own gapless excitations. Nevertheless, for every quantized vortex added to the system, there emerges a new (usually gapped) low-energy branch of Kelvin--Tkachenko quasiparticle excitations \cite{Simula2013a}. 

Here we study the zero modes of rotating Bose--Einstein condensates by computationally solving the quasiparticle spectra of a quantum degenerate Bose gas, which is exposed to an external vector potential that causes the atoms to rotate at an angular velocity $\Omega$ around the chosen $z$-axis of the three-dimensional harmonic potential. Surprisingly, by applying the microscopic Bogoliubov--deGennes field theory, we find this symmetry broken system to be capable of hosting \emph{two zero modes}. The first one is the usual ``trivial" Bogoliubov phonon and the second one is a Tkachenko vortex wave. We discuss the properties of the Tkachenko zero mode and its experimental signatures.

\section{Model}
By using the method of second quantization, Bogoliubov provided an explanation of the phenomenon of superfluidity in terms of a gas of quasiparticles \cite{Bogoliubov1946a}. In this Bogoliubov quasiparticle picture, the boson field 
$\hat{\Psi}({\bf r}) = \hat{\phi}_0({\bf r}) +\hat{\psi}^\dagger({\bf r})$ is expressed in terms of the macroscopically occupied condensate $\hat{\phi}_0({\bf r})$, which is treated as a classical field, and the fluctuation operator $\hat{\psi}^\dagger({\bf r})$ \cite{Bogoliubov1946a,Fetter1972a,Lewenstein1996a}. By introducing the quasiparticle annihilation $\eta$ and creation operators \cite{Lewenstein1996a}
\begin{equation}
\eta^\dagger = \int\left(u_q({\bf r}) \hat{\psi}({\bf r}) + v_q({\bf r}) \hat{\psi}^\dagger({\bf r}) \right)d {\bf r},
\label{eta}
\end{equation}
where $u_q({\bf r})$ and $v_q({\bf r})$ are the particle and hole quasiparticle amplitudes $(u_q({\bf r}), v_q({\bf r}))$ of the Bogoliubov spinor the corresponding eigenenergies $E_q$ of the quasiparticle states can be explicitly calculated for a given physical system by solving the Bogoliubov--deGennes equations \cite{Bogoliubov1946a,deGennes1964a,Fetter1972a}
\begin{eqnarray}
(\mathcal{L}({\bf r})-\mu)u_q({\bf r}) - \mathcal{M}({\bf r})v_q({\bf r})  &=&  E_q u_q({\bf r})\\
(\mathcal{L^*}({\bf r})-\mu)v_q({\bf r}) - e^{i\theta}\mathcal{M^*}({\bf r})u_q({\bf r})  &=& -E_q v_q({\bf r})
\label{BDG}
\end{eqnarray}
where $\mathcal{L}({\bf r})$ and $\mathcal{M}({\bf r})$ are model dependent operators and $\mu$ is the chemical potential. For the boson system studied here $\theta=0$, whereas for fermion systems the anticommutation relations yield $\theta=\pi$. The  mixing of the particle and hole character in the quasiparticle operators $\eta$ and $\eta^\dagger$ yields a pseudo spin half structure even for the bosonic quasiparticles. For systems whose finite-energy excitations are fermionic the excitation energies $E_q$ are measured with respect to the Fermi energy $E_{\rm F}$, whereas for boson systems they are measured with respect to the chemical potential $\mu$ of the condensate. 

For a bosonic system the Eqns (2) and (3) have a ``trivial" zero energy solution $E_q=0$ with $u_q({\bf r})=v^*_q({\bf r})$. Inserting this solution to Eq.(\ref{eta}), shows that such zero energy quasiparticles satisfy the relation $\hat{\eta}_q = \hat{\eta}^\dagger_q$ , which is the defining property of a Majorana operator---a (quasi) particle creation operator is equivalent to its annihilation operator. Since such zero modes satisfy neither boson commutation or fermion anticommutation relations \cite{Stone2006a}, their character is that of an anyon \cite{Wilczek1982a}.

\begin{figure}
\includegraphics[width=0.5\columnwidth]{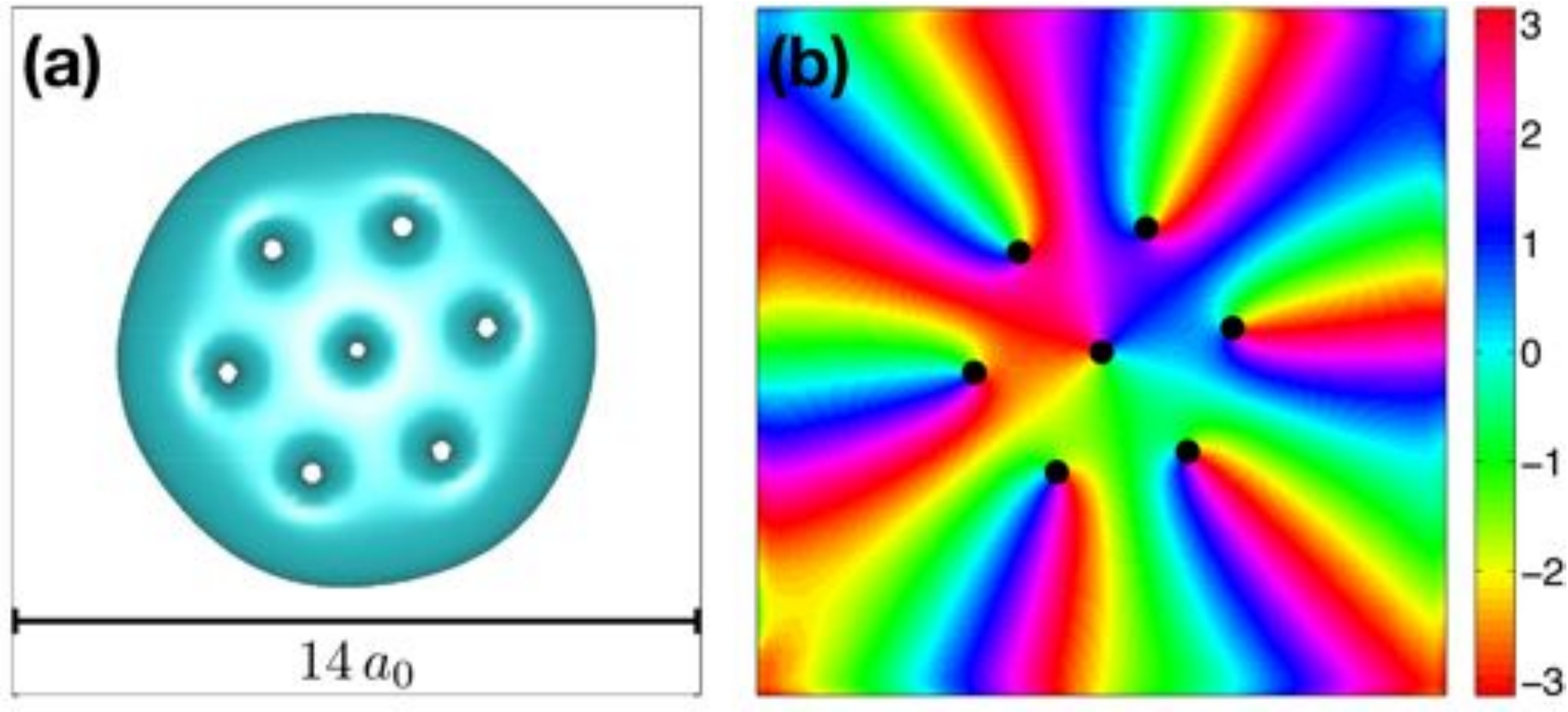}
\caption{First $q=1$, zero mode $\phi({\bf r})$ of a condensed Bose gas rotating at an angular frequency $\Omega=0.58\omega_\perp$. (a) Isosurface of the condensate density $|\phi({\bf r})|^2$ with seven quantized vortices piercing the condensate forming a sixfold symmetric triangular array, viewed along the rotation axis. (b) The corresponding phase map $S({\bf r})=\arg(\phi({\bf r}))$ shows the seven phase singularities at the vortex cores, marked with filled circles.} 
\label{fig1}
\end{figure}

\begin{figure}
\includegraphics[width=0.5\columnwidth]{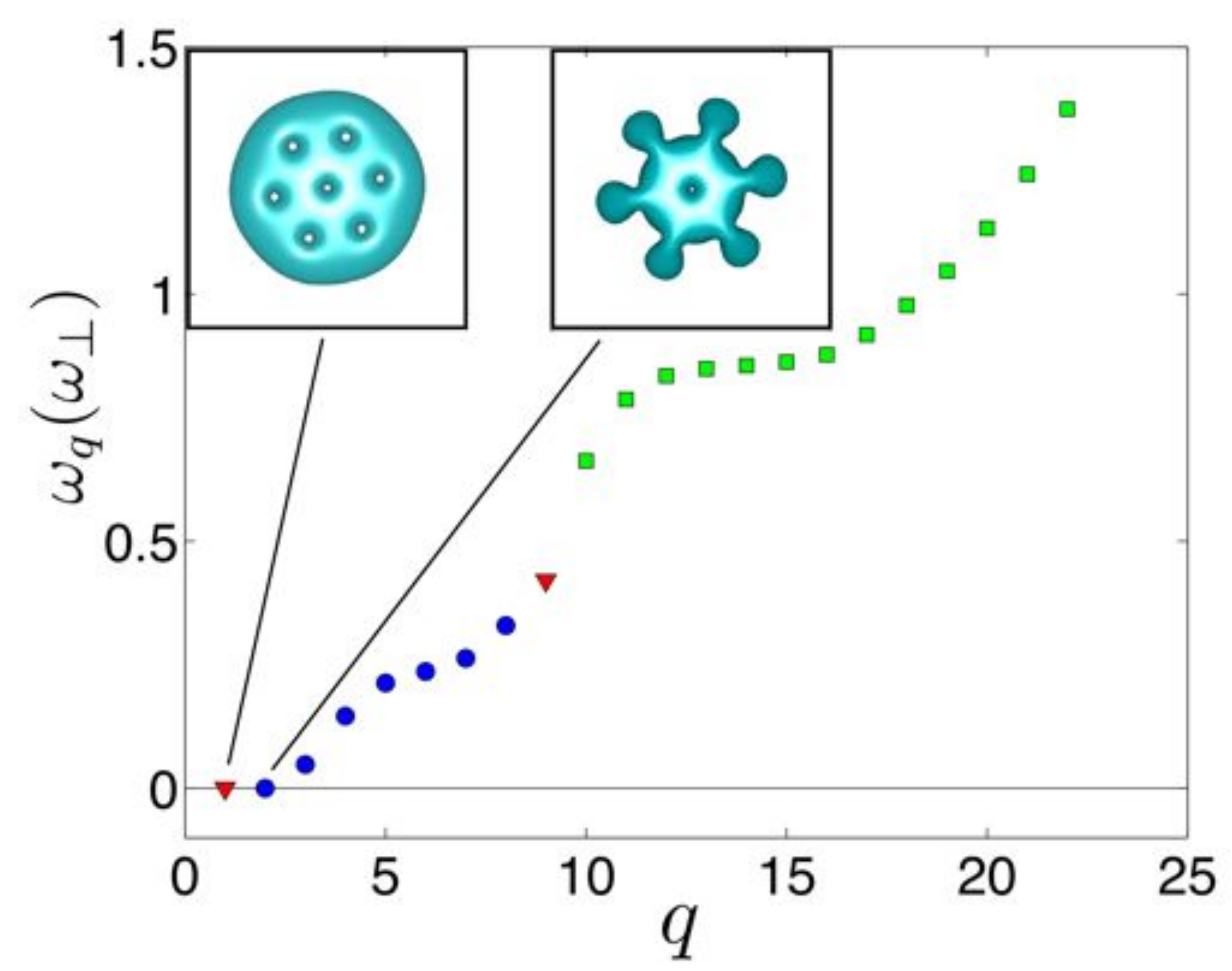}
\caption{Quasiparticle excitation frequencies $\omega_q$ as a function of an integer index $q$ which labels the modes in the order of increasing energy. The two zero modes are singled out in the insets. The lowest compressible Bogoliubov sound modes ($q=1$ and $q=9$ Kohn mode) are marked with triangles. The seven vortex displacement modes corresponding to the seven vortex degrees of freedom are labeled by circles and the remaining square markers correspond to surface waves.} 
\label{fig2}
\end{figure}

We consider a Bose--Einstein condensate of approximately $N=3\times10^4$ atoms at zero temperature limit. The $^{87}$Rb atoms of mass $m$ are confined in a three-dimensional harmonic oscillator potential $V({\bf r})=m(\omega^2_\perp r^2_\perp + \omega^2_zz^2)/2$, which rotates at an angular frequency $\Omega=0.58\omega_\perp$ unless otherwise stated. This rotation frequency stabilizes the seven vortex array as the ground state of this system. The interaction coupling constant $g=4\pi\hbar^2a/m$, where $a=4.76$nm is the $s$-wave scattering length of the atoms, and the oscillator length $a_0=\sqrt{\hbar/m\omega_\perp}$. We have chosen $\omega_\perp=\omega_z/\sqrt{8}=2\pi\times100 \,{\rm Hz}$ and the interaction parameter $gN=1000 \,\hbar\omega_\perp a_0^3$. For this system, the operators in Eq.~(\ref{BDG}) are $\mathcal{M}({\bf r})=g\phi^2({\bf r})$ and $\mathcal{L}({\bf r})= -\frac{\hbar^2\nabla^2}{2m}+V({\bf r})+2g|\phi({\bf r})|^2  - \Omega L_z$, where $L_z$ denotes the projection of the orbital angular momentum operator onto the axis of rotation. The condensate order parameter $\phi({\bf r})$ satisfies the Gross--Pitaevskii equation \cite{Pitaevskii1961a,Gross1961a}
\begin{equation}
\left(\mathcal{L}({\bf r})-\mu-g|\phi({\bf r})|^2 \right) \phi({\bf r}) = 0.
\label{GP}
\end{equation}
To draw an analogy with charged superconductors, the operator $\mathcal{L}({\bf r})$ may alternatively be expressed in terms of a vector potential ${\bf A}_\Omega= m^2\Omega( - y{\bf e}_x + x{\bf e}_y )/h$ as
$\mathcal{L}({\bf r})=[(i\hbar\nabla -  \kappa {\bf A}_\Omega)^2 +m^2(\omega^2_\perp-\Omega^2) r_\perp^2+m^2\omega^2_z z^2 +4mg|\phi({\bf r})|^2]/2m$, where the vortex charge $\kappa=h/m$ and $h$ is Planck's constant.

The coupled equations (\ref{BDG}) possess a particle-hole symmetry such that given a solution $u_q({\bf r}),v_q({\bf r}),E_q$ a second solution $u'_q({\bf r}),v'_q({\bf r}),E'_q$ always exists and is obtained by a simple transformation  
$u'_q({\bf r})=v^*_q({\bf r})$, $v'_q({\bf r})=u^*_q({\bf r})$ and $E'_q=-E^*_q$. For fermion systems both hole and particle type excitations are treated on an equal footing whereas for bosonic systems half of all excitations have a negative norm and in order to avoid describing them as fermionic excitations, they are typically discarded. In addition, for a Bose--Einstein condensed system, there exists a trivial solution, $E_1=0$ to Eq.~(\ref{BDG}), which is obtained by setting $u_1({\bf r})=v^*_1({\bf r})=\phi({\bf r})$. This is the zero mode related to the spontaneous symmetry breaking associated with $U(1)$ gauge symmetry and the formation of the phase coherent Bose--Einstein condensate. Such a zero mode does not obey the bosonic quasiparticle orthonormalization condition \cite{Fetter1972a}
$
\int (u^*_q({\bf r}) u_{p}({\bf r}) -v^*_{q}({\bf r}) v_{p}({\bf r}) )d^3{\bf r}=\delta_{qp} 
$
and
$
\int (u_q({\bf r}) v_{p}({\bf r}) -u_{p}({\bf r}) v_{q}({\bf r}) )d^3{\bf r}=0,
$
instead satisfying the relation  
$
\int u^*_1({\bf r}) u_{1}({\bf r}) d^3{\bf r}  = \int v^*_{1}({\bf r}) v_{1}({\bf r}) d^3{\bf r}. 
$
In addition to the zero energy solution  $u({\bf r})=\phi({\bf r})$, other zero modes degenerate with the ground state for which $|u_q({\bf r})|\ne |\phi({\bf r})|$ and $
(\mathcal{L}({\bf r})-\mu) u_q({\bf r}) = \mathcal{M}({\bf r})u^*_q({\bf r}),
$
may exist in the spectrum as discussed in the following.

\section{Results}
We have calculated the quasiparticle excitation spectra by numerically solving Eq.~(\ref{BDG}) using a parallelized Arnoldi iteration method \cite{Virtanen2001a,Virtanen2001b}. The three dimensional operators are discretized using a finite-element discrete variable representation to yield a sparse matrix representation for the Bogoliubov--de Gennes operator \cite{Schneider2006a,Simula2008b,Simula2008a,Simula2010a}. Before solving the Bogoliubov--de Gennes problem, the condensate ground state $\phi({\bf r})$ and the corresponding chemical potential $\mu$ are obtained by solving the Gross--Pitaevskii equation (\ref{GP}) using an over-relaxation method with the chemical potential computed to an accuracy better than six decimal places. The results presented here are from the same data set discussed in \cite{Simula2013a}.

Figure \ref{fig1}(a) shows a condensate density $|\phi({\bf r})|^2$ isosurface and (b) the corresponding phase map $S({\bf r})$ for the calculated ground state wavefunction $\phi({\bf r})=|\phi({\bf r})|\exp(iS({\bf r}))$. The chemical potential of this condensate is $\mu=\langle\phi| \mathcal{L}({\bf r})-g|\phi({\bf r})|^2|\phi\rangle=12.0 \,\hbar\omega_\perp$ and the orbital angular momentum $L=\langle\phi| -i\hbar (x\partial_y-y\partial_x)|\phi\rangle=4.2 \,\hbar N$. The condensate is rotated at angular frequency $\Omega=0.58\omega_\perp$ and is pierced by seven singly quantized vortex filaments. The circulation $\oint v_s({\bf r}) \cdot dl=\kappa \ell$ of the superfluid velocity $v_s({\bf r})=\hbar/m\nabla S({\bf r})$ around the vortex cores is quantized in integer $\ell$ multiples of $\kappa=h/m$, as predicted by Onsager. In the ground state of the rotating potential, these vortices are arranged in a triangular lattice structure with sixfold discrete rotational symmetry. The phase singularities located inside the vortex cores are marked with filled circles in Fig. \ref{fig1}(b).


Figure \ref{fig2} shows the Bogoliubov excitations calculated for the rotating condensate ground state shown in Fig.~\ref{fig1}. Frequencies of the lowest collective quasiparticle modes are shown as a function of an integer index $q$, which labels the quasiparticles in increasing order of their energy. The most prominent feature of this excitation spectrum is the presence of \emph{two zero modes} labelled by $q=1$ and $q=2$. We note that all shown states are duplicated in the numerically calculated spectrum due to the aforementioned particle-hole symmetry of the Bogoliubov--de Gennes equations and thus we find altogether four zero modes, only half of which represent distinct quasiparticles and which are shown in Fig.~\ref{fig2}. The first one $q=1$ of these modes $\phi({\bf r})$ is always present in these Bose--Einstein condensed systems irrespective of the value of the vector potential ${\bf A}_\Omega$. It corresponds to the Bose--Einstein condensate of atoms and is equivalent to the state shown in Fig.~\ref{fig1}. The presence of this zero mode is the consequence of a spontaneous symmetry breaking associated with the formation of the Bose--Einstein condensate. 

\begin{figure}
\includegraphics[width=0.5\columnwidth]{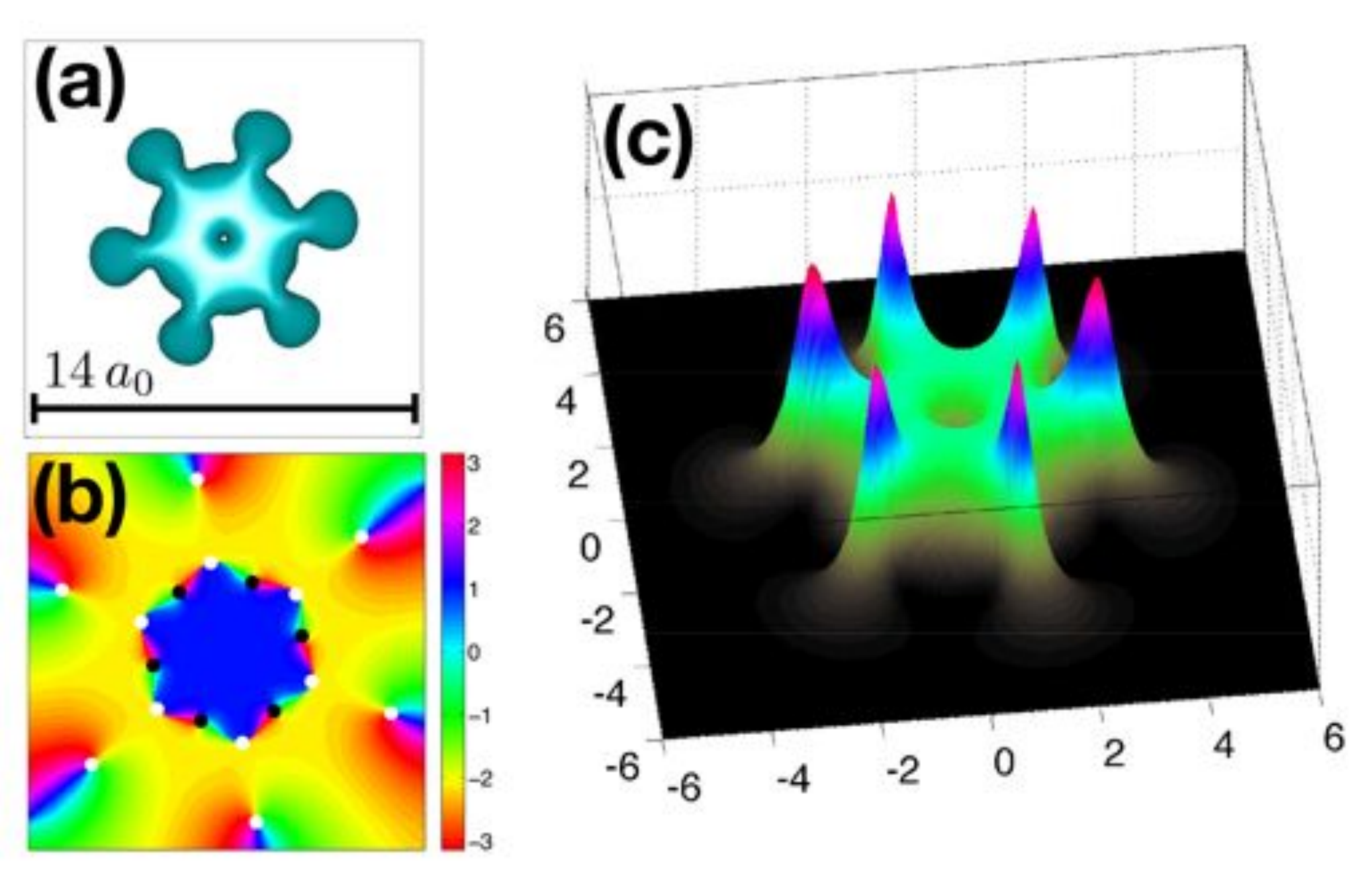}
\caption{Second $q=2$ zero mode $\psi({\bf r})$ of a condensed Bose gas rotating at an angular frequency $\Omega=0.58\omega_\perp$. (a) Isosurface of the density $|\psi({\bf r})|^2$ viewed along the rotation axis. (b) Relative phase map $M({\bf r})=\arg(\phi({\bf r}))-\arg(\psi({\bf r}))$ shows 18 phase singularities. The six outermost of these are ghost vortices of $\psi({\bf r})$ and reside at the edge of the condensate. The remaining 12 phase defects correspond to 6 vortices in the condensate ground state $\phi({\bf r})$ and 6 antivortices in the state $\psi({\bf r})$. Both $\phi({\bf r})$ and $\psi({\bf r})$ have a vortex in the center of the trap but these vortices are of opposite circulation and cancel each other in the relative phase map. Frame (c) shows quasiparticle probability density of the $q=2$ zero mode $\psi({\bf r})$ in the $z=0$ plane. The six peaks are localized in the vortex cores of the ground state $\phi({\bf r})$. At the origin, there is a phase singularity in $\psi({\bf r})$ where the probability density vanishes.} 
\label{fig3}
\end{figure}

For this seven vortex system we observe a second $q=2$ zero mode $\psi(\bf r)$. This is related to the fact that in the rotating condensate ground state the continuous rotation symmetry SO(2) has also been broken and only the reduced discrete sixfold rotational symmetry of the vortex lattice remains. The source of this symmetry breaking can be traced back to the vector potential ${\bf A}_\Omega$ in the Hamiltonian. We emphasize that this $q=2$ zero mode is distinctly different from the high angular momentum surface mode, which approaches zero energy when new vortices are nucleated in the system \cite{Simula2002a,Ueda2006a,Dagnino2009a}.  Fig.~\ref{fig3}(a) shows a density isosurface plot $|\psi({\bf r})|^2=|u_2({\bf r})|^2=|v_2({\bf r})|^2$ viewed along the rotation axis and Fig.~\ref{fig3}(b) shows a relative phase map  $M({\bf r})=\arg(\phi({\bf r}))-\arg(\psi({\bf r}))$ for the $q=2$ zero mode. The white markers denote the locations of vortex cores in the state $\psi(\bf r)$ and the black markers correspond to the vortex phase singularities in the condensate ground state. Both $\phi(\bf r)$ and $\psi(\bf r)$ have a vortex phase singularity at the origin, which cancel each other in the relative phase map shown in Fig.~\ref{fig3}(b). Figure.~\ref{fig3}(c) shows the probability density of the $q=2$ mode in the $z=0$ plane. It can be viewed as a bound state of the effective potential created by the combination of the harmonic trap and the condensate mean-field. There are six prominent peaks which are localized inside the vortex cores of the condensate. There are also seven vortex phase singularities in $\psi(\bf r)$. Six of these are located in between the peaks observed in Fig. \ref{fig3}(c) and the seventh is located on the $z$-axis in the center of the figure. 

The Kohn mode or the center-of-mass dipole mode with frequency $\omega_\perp-\Omega$ has been marked in Fig.~\ref{fig2} by a triangle. The two insets in Fig.~\ref{fig2} show density isosurface plots of the two zero energy states $\phi(\bf r)$ and $\psi(\bf r)$ with $q=1$ and $q=2$, respectively. The states with $q=2-8$ are the lowest excitation modes in each of the seven Kelvin--Tkachenko vortex wave branches present in this system due to the seven vortex degrees of freedom \cite{Simula2013a,Simula2010a}.

\begin{figure}
\includegraphics[width=1\columnwidth]{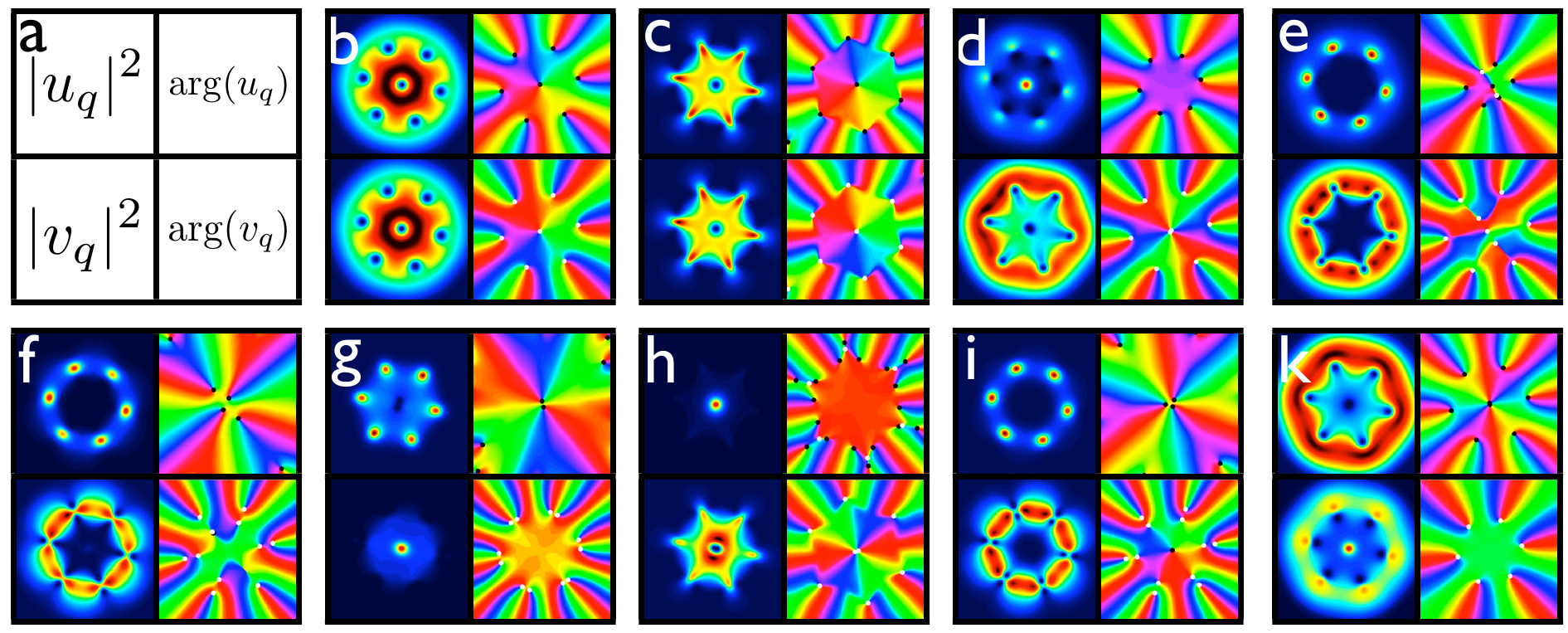}
\caption{Probability densities and phase maps of the $q=1-9$ quasiparticle amplitudes $\left( u_q({\bf r}),v_q({\bf r}) \right)$ of the Bogoliubov spinors. Frame (a) is a notation guide for the frames (b)-(k). Frames (b) and (c) show the $q=1$ and $q=2$ zero modes, respectively. Frames (c)-(i) are the seven low-lying Kelvin--Tkachenko modes \cite{Simula2013a} and (k) shows the centre of mass Kohn mode. Note in particular that the probabilities $|u_1({\bf r})|^2 = |v_1({\bf r})|^2$ and $|u_2({\bf r})|^2 = |v_2({\bf r})|^2$ which distinguishes these zero modes from all other quasiparticles in the system. Also note the time reversal symmetry between the particle and hole components visible in the phase maps of the two zero modes. 
} 
\label{fig4}
\end{figure}

Figure \ref{fig4} shows the particle $u_q(\bf r)$ and hole $v_q(\bf r)$ probability amplitudes of the Bogoliubov spinors and their corresponding phase maps for the $q=1-9$ quasiparticles. The location and sign of the circulation of the quantized vortices are marked in the phase plots with filled circles. Frame (a) is a guide for notation. Frame (b) corresponds to the $q=1$ zero mode and frame (k) shows the centre-of-mass Kohn mode. Frames (c)-(i) are the lowest modes in the seven Kelvin--Tkachenko mode branches corresponding to the seven vortex degrees of freedom \cite{Simula2013a}. The $q=1$ (b) and $q=2$ (c) zero modes are qualitatively distinct from all other quasiparticles in the spectrum. As seen in (b) and (c) the particle and hole components of the zero modes are equal $|u_1({\bf r})|^2 = |v_1({\bf r})|^2$ and $|u_2({\bf r})|^2 = |v_2({\bf r})|^2$, which results in their anomalous normalization condition and reflects their Majorana property $\hat{\eta} = \hat{\eta}^\dagger$. Note also that the phase maps reveal the particle and hole amplitudes to be time-reversal symmetric partners. Interestingly, also the angular momenta of the $q=1$ and $q=2$ states are equal. In the Feynman interpretation of antiparticles \cite{Stueckelberg1941a,Feynman1949a}, a zero mode for which a particle and its antiparticle are equivalent could be described in terms of simultaneous forward and backward time propagation of an excitation. Naively applying this reasoning to the quasiparticle zero mode in our system and noting that time-reversal changes the sense of rotation, we may characterize the zero energy Tkachenko mode as a quasiparticle which comprises of simultaneous left and right handed rotation of the vortex lattice.

The self-consistency of the calculated spectra are guaranteed by the fact that we obtain the $E_1=\hbar\omega_1=0$ Nambu--Goldstone solutions for all calculated ground states with a numerical precision $\rm{Re}(\omega_1/\omega_\perp)<10^{-9}$, $\rm{Im}(\omega_1/\omega_\perp)<10^{-3}$. The $q=2$ zero mode which appears for the seven vortex system the corresponding accuracies are $\rm{Re}(\omega_2/\omega_\perp)<10^{-10}$, $\rm{Im}(\omega_2/\omega_\perp)<2\times10^{-2}$. In addition, all calculated $q=1$ zero modes obey the normalization condition, $
\int |u_{1}({\bf r})|^2 d^3{\bf r}  / \int |v_{1}({\bf r})| d^3{\bf r}-1 <10^{-6}$ and for the $q=2$ zero mode we obtain $
\int |u_{2}({\bf r})|^2 d^3{\bf r}  / \int |v_{2}({\bf r})| d^3{\bf r}-1 <10^{-7}$. This normalization property qualitatively distinguishes these modes from all other, non-zero, modes in the spectra, see Fig.~\ref{fig4}.

Although in calculations where the Bogoliubov diagonalization is performed for a condensate in an excited metastable state or if the condensate exists at the boundary of a stability window with respect to some external parameter, it is not unusual to find excitations with imaginary frequencies, which in those cases are interpreted to signal a dynamical instability of the condensate. However, in the problem considered here the presence of the nonzero imaginary parts is attributed to the numerical precision when inverting a matrix with a (near) singular spectrum iteratively instead of a dynamical instability pertaining to the physical system. This is because here the diagonalization is performed for a \emph{stable ground state} condensate which, by definition, means that there does not exist a state to which the condensate could transition via a dynamical instability. Furthermore, the rotation frequencies are chosen in such way that they lie well within the stability window of the given number of vortices in the system. For example we obtain seven vortices in a ground state also for $\Omega=0.56\omega_\perp$ and $\Omega=0.60\omega_\perp$. Further evidence for the stability of the rotating ground state in comparison to a state with different number of vortices is provided by the fact that there are no surface modes localized near the condensate boundary, which appear in systems in which more vortices are about to nucleate \cite{Simula2002a}. Moreover, for non-ground state condensates which exhibit dynamical instabilities, such as a doubly quantized vortex state, there is always a clear reason for the instablity to emerge, such as the splitting of the double charge vortex in to the energetically more favourable state of two single quantum vortices \cite{Simula2002b}. We are not aware of any such destabilizing mechanism, which would be able to induce a dynamical instability for the Tkachenko mode in these slowly rotating systems.


The physical character of the quasiparticle state $\psi({\bf r})$ is that of a Tkachenko vortex wave \cite{Tkachenko1966a}. It comprises of elliptically polarized collective motion of the vortices around their equilibrium locations \cite{Rajagopal1964a,Tkachenko1966a,Fetter1975a,Williams1976a,Sonin1976a,Baym1983a,Sonin1987a,Baym2003a,Gifford2004a}. This mode has transverse and longitudinal polarization components corresponding to azimuthal and radial motion of the vortices. The Tkachenko modes of vortex lattices in harmonically trapped Bose--Einstein condensates have been calculated analytically within a vortex continuum approximation and the mode frequencies have been found to approach zero in the centrifugal limit $\Omega\to\omega_\perp$ as the system approaches the quantum-Hall regime where the vortex lattice is predicted to melt \cite{Baym2003a,Snoek2006a}. This is in agreement with the experimental observations for large vortex lattices \cite{Coddington2003a,Schweikhard2004a}. However, the Tkachenko mode has not been predicted to appear at zero energy for systems with nonzero $\Omega<\omega_\perp$. Moreover, previous numerical calculations for purely two-dimensional systems \cite{Simula2004a,Mizushima2004a,Woo2004a,Baksmaty2004a} or three-dimensional prolate systems \cite{Simula2010a} have not predicted a zero mode other than the $q=1$ condensate mode in these systems.

\begin{figure}
\includegraphics[width=0.5\columnwidth]{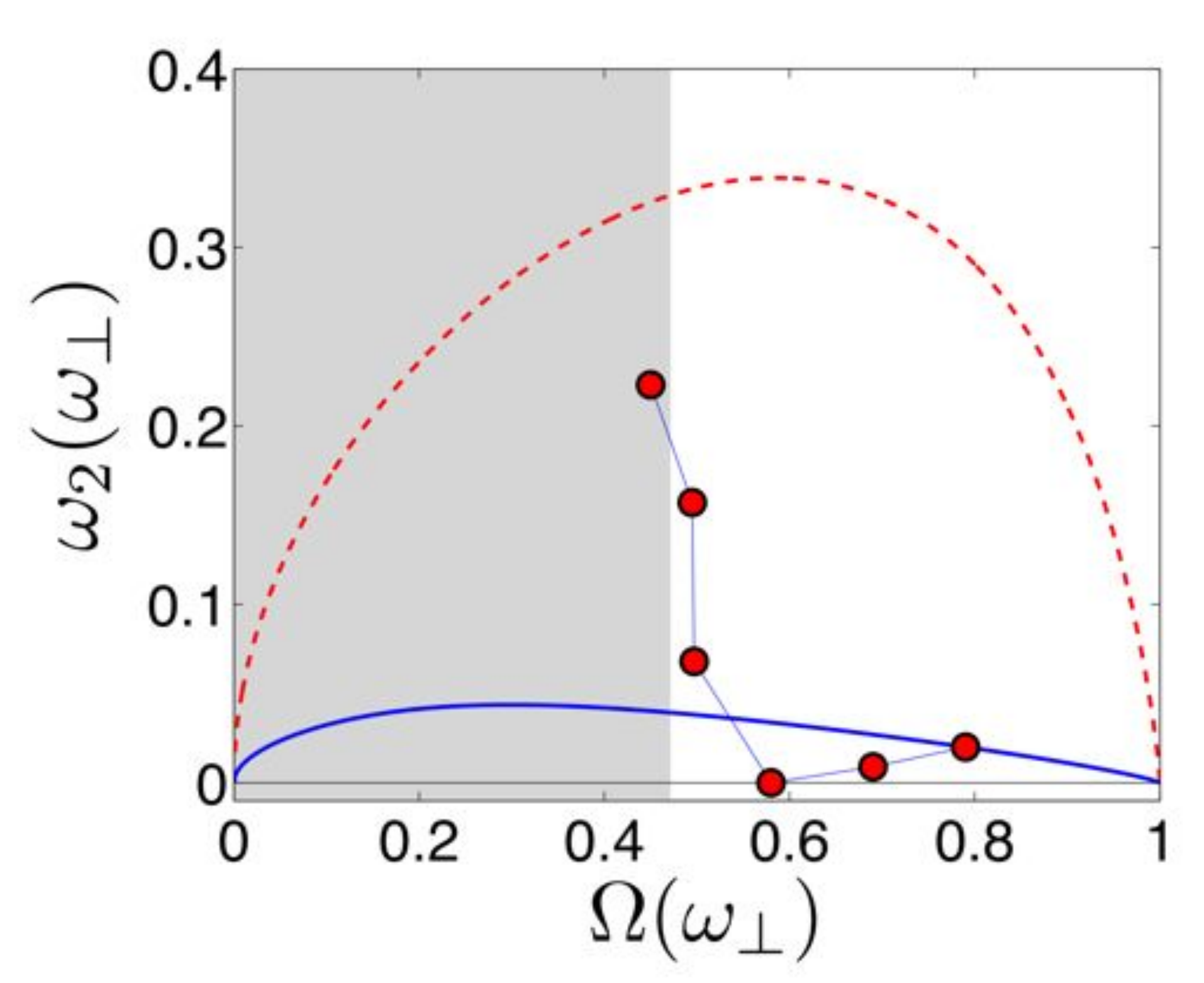}
\caption{Frequency $\omega_2$ of the $q=2$ Tkachenko mode as a function of $\Omega$ corresponding to 1, 2, 3, 7, 12, and 19 vortex ground states (filled circles) and the dashed and solid curves are continuum theory predictions using two different values, $\alpha = 5.45$ and $\alpha=1.08$ respectively, for the effective wavevector $\alpha$, see Eq.~(19) in \cite{Baym2003a}. The gray shaded area corresponds to vortex free region in the Thomas--Fermi approximation.} 
\label{fig5}
\end{figure}

We have performed calculations for $\Omega/\omega_\perp$ in the range 0-0.8 for the present three-dimensional oblate system.
Figure~\ref{fig5} shows the lowest Tkachenko mode (filled circles) as a function the angular rotation frequency of the trap corresponding to 1, 2, 3, 7, 12 and 19 vortices in the ground state, see also \cite{Simula2013a}. The minimum seen in this curve reaches zero for the case of seven vortices. For comparison, we have also plotted two theory predictions obtained using two different values, $\alpha = 5.45$ and $\alpha=1.08$ respectively, for the effective wavevector $\alpha$ in Eq.~(19) of Ref.~\cite{Baym2003a}. The assumption of the hydrodynamic continuum theories that the vortex core size is small in comprarison to the intervortex distance in the lattice is not satisfied by these small vortex arrays which explains the disagreement between the calculated Bogoliubov modes and the continuum theory predictions. Furthermore, the shaded gray area corresponds to a region below the critical vortex nucleation frequency within the Thomas--Fermi approximation, and does not contain any vortices and therefore also the Tkachenko mode does not exist in that region. It is thus clear that the continuum theory is not applicable for these finite-size few vortex arrays and therefore further work is required in order to theoretically explain the low-lying dispersion of Tkachenko modes.

The observation of a minimum in the Tkachenko mode frequency, which reaches zero energy at intermediate rotation frequencies is surprising since it is not predicted by hydrodynamic continuum equations. Similar minimum in the Tkachenko mode frequency, plotted as a function of the trap rotation frequency, was also observed in a prolate three dimensional system where the minimum was observed at nonzero excitation energy \cite{Simula2010a}. That seven classical vortices in a ring arrangement is a marginally stable configuration having a normal mode with precisely zero frequency \cite{Thomson1883a,Havelock1931a} is therefore likely a coincidence rather than a generic property of seven vortex systems. 

\begin{figure}
\includegraphics[width=0.5\columnwidth]{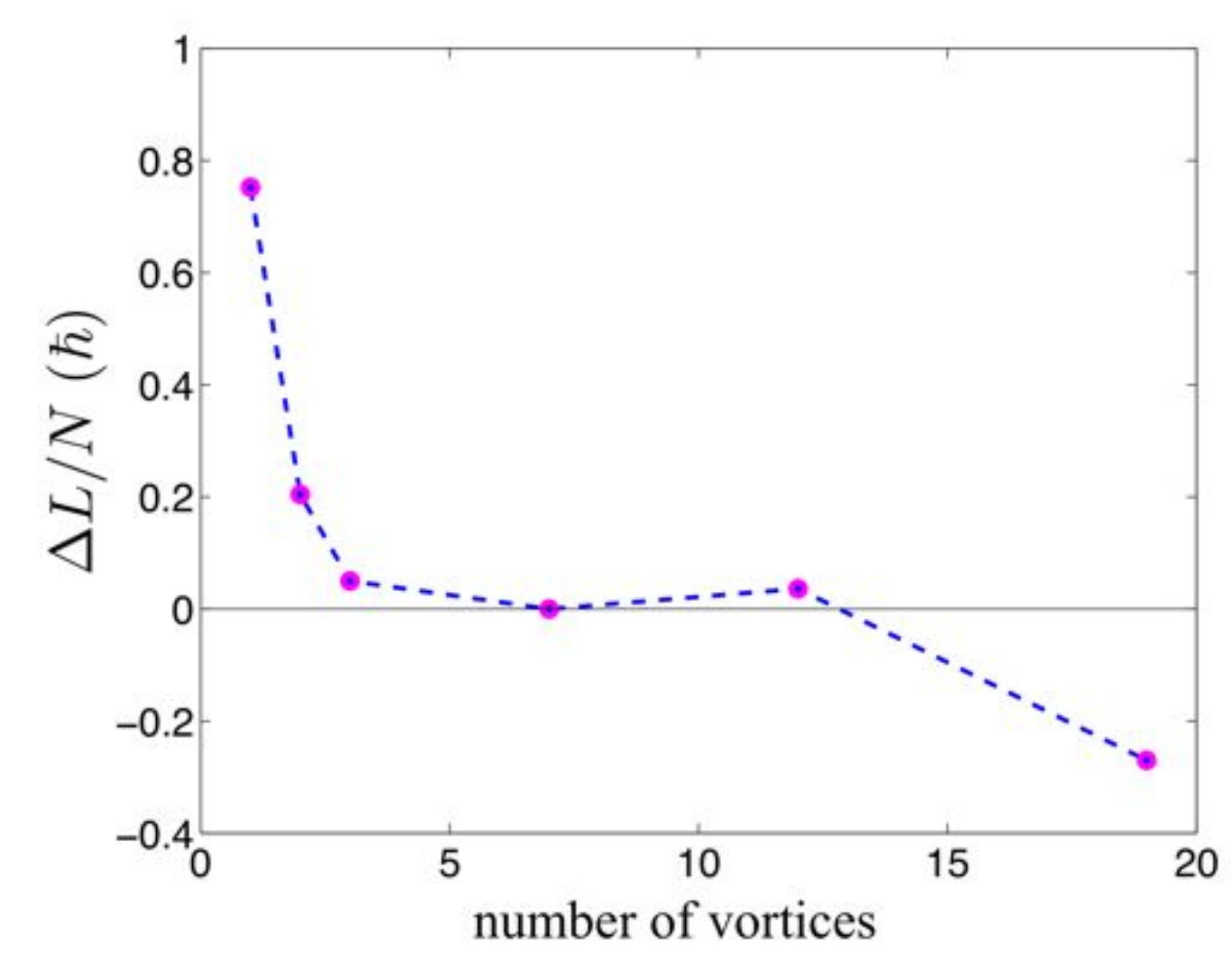}
\caption{Difference $\Delta L =(L_2-L_1)/N$ in angular momentum per particle of the two respective zero modes $q=1$ (condensate ground state) and $q=2$ (Tkachenko mode) as a function of the number of vortices in the condensate. The second zero mode emerges where this curve crosses zero where the angular momentum of the Tkachenko mode becomes equal to that of the condensate.} 
\label{fig6}
\end{figure}

Further clues for why such minimum in the Tkachenko mode emerges may be obtained by considering the angular momenta of the Tkachenko mode quasiparticles. Figure \ref{fig6} shows the difference $\Delta L=(L_2-L_1)/N$ between the angular momentum per particle of the respective two lowest excitation modes in the system plotted as a function of number of vortices in the rotating ground state corresponding to different trap rotation frequencies. It shows that the difference $\Delta L=(L_2-L_1)/N$ changes sign having a zero crossing for small number of vortices in the system. This observation suggests the following interpretation. Rotation of the condensate atoms and vortices are coupled by Magnus force, which drives the vortex dynamics. Therefore the vortices cannot move in the condensate independently of the condensate superflow. This coupling between the superfluid and its vortices results in the finite frequency for the Tkachenko modes. For zero energy Tkachenko mode the ``rotational symmetry breaking is restored" where the angular momentum of the vortex field is equal to the angular momentum of the condensate field and the coupling between these two fields vanishes making them noninteracting. Hence the vortex lattice can execute purely azimuthal rotation without affecting the atom field. 

For nonzero Tkachenko frequencies, the angular rotation frequency of the excited vortex lattice is different from that of the condensate. Therefore, if the Tkachenko mode is populated it will cause the vortex lattice to rotate faster than the equilibrium lattice. This will increase the vortex density $n_v\propto \Omega$. The increase in vortex density would increase angular momentum but since angular momentum is conserved the condensate atoms must be rearranged, which causes superflow that couples back to the motion of vortices via the Magnus force. This self-consistent coupling between the atoms and vortices together with angular momentum conservation results in the elliptically polarized motion of the vortices in the Tkachenko modes and their nonzero oscillation frequency. In a three dimensional system the third axial dimension provides an additional degree of freedom which can be used to``deposit" angular momentum. This means that it is possible to find parameters for which the angular momentum of the system becomes fine tuned with the vortex lattice so that there is no differential rotation between the lattice and the condensate due to the Tkachenko mode. Figure \ref{fig6} shows that for trap rotation frequencies below the Tkachenko mode minimum the angular momentum of the Tkachenko mode is larger than the condensate and therefore the vortex field tries to speed up the condensate. For trap rotation frequencies well above the Tkachenko mode minimum the angular momentum of the Tkachenko mode is smaller than the condensate and the vortex field tries to slow down the condensate. At zero Tkachenko frequency there is no tension between the superfluid and the vortex lattice.

Finally, we note that in a classical unbounded point vortex system, there are two linearly dependent eigenmodes with orthogonal polarizations \cite{Campbell1981a}. The other corresponds to pure azimuthal rotation and has zero energy. The other corresponds to breathing or purely radial/longitudinal motion of the vortices. These two behaviour correspond to two polarization extremes of the same elliptical vortex trajectories. In a Bose--Einstein condensate with sufficiently strong dissipation the vortices in a lattice state will indeed spiral radially out of the system. In a Bose--Einstein condensate with zero energy Tkachenko mode and zero dissipation the vortex motion will be purely azimuthal.

\section{Conclusions}

In conclusion, we have identified a new, slowly rotating, regime of harmonically trapped Bose--Einstein condensates by observing a minimum in the Tkachenko mode calculated as function of the angular rotation frequency of the external trapping potential. The observed minimum, which occurs in a few-vortex system does not exist in the theoretical predictions based on continuum hydrodynamics. We have found such a minimum to harbour a zero energy quasiparticle state in the elementary excitation spectrum of a rotating Bose--Einstein condensate calculated for a system in a three-dimensional harmonic trap hosting seven singly quantized vortices. The physical character of this zero mode is that of a Tkachenko vortex wave. The zero energy Tkachenko mode has the exceptional character of corresponding to a purely azimuthal motion of the vortices with zero amplitude in the radial component. As opposed to this, the lowest Tkachenko mode generically has a finite excitation frequency consistent with the presence of a radial component of the associated vortex motion. 

Since there is no energy cost associated with populating a zero energy Tkachenko excitation mode, its presence implies the \emph{possibility} of formation of a quasiparticle Bose--Einstein condensate of quanta of Tkachenko waves. However, since the equations used here are linearized around the macroscopically occupied rotating condensate ground state, based on these calculations only, it is not possible to judge whether such a condensate will form in these systems or whether the degeneracy of the two zero modes will be lifted if the $q=2$ mode is populated appreciably. The double degeneracy of the ground state of this system implies a finite entropy of the zero temperature ground state. For $\Omega=\omega_\perp$ quantum Hall effects and the associated topological insulators are expected to appear in rotating quantum gases with a massively degenerate ground state. Here we observe ground state degeneracy already at finite $\omega_\perp-\Omega$. However, there remains an important question regarding the robustness of this zero mode with respect to quantum and thermal fluctuations and perturbations in external parameters. 

These results show that the existence of Nambu--Goldstone zero modes associated with spontaneously broken symmetries can depend on the subtle details of the system specific parameters. The predicted minimum in the Tkachenko mode could be experimentally observed by the virtue of an absence of torsional vortex lattice vibration. The nonzero Tkachenko modes can be excited in rapidly rotating Bose--Einstein condensates by inducing small perturbations to the equilibrium vortex lattice using external potentials \cite{Coddington2003a,Schweikhard2004a}. Similar technique to excite Tkachenko modes could also be applied to slowly rotating few-vortex condensates and the resulting torsional vortex motion detected using quasi-in-situ imaging of vortex dynamics \cite{Navarro2013a}. Using such methods it should be possible to establish the presence of a minimum in the Tkachenko mode at intermediate rotation frequencies, however, observing the zero mode directly would be difficult using this technique since it would take infinitely long to measure the period of a zero frequency excitation corresponding to (non)motion of the vortices.

\section{Acknowledgements}
I am grateful to Michael Morgan for valuable discussions.


\end{document}